\documentclass[prd, aps, superscriptaddress, preprintnumbers, twocolumn, floatfix, nofootinbib]{revtex4}
\pdfoutput=1

\usepackage{amsfonts}
\usepackage{amsmath}
\usepackage{amssymb}
\usepackage{bm}
\usepackage{dcolumn}
\usepackage{graphicx}   
\usepackage[latin1]{inputenc}
\usepackage{latexsym}
\usepackage{rotating}
\usepackage{hyperref}
\usepackage{graphicx}
\usepackage{color}

\newcommand\be{\begin{equation}}
\newcommand\ba{\begin{eqnarray}}
\newcommand\ee{\end{equation}}
\newcommand\ea{\end{eqnarray}}

\newcommand{\dd}{{\rm d}}

\begin{document}

\title{Strengthening the TCC Bound on Inflationary Cosmology}

\author{Robert Brandenberger}
\email{rhb@physics.mcgill.ca}
\affiliation{Department of Physics, McGill University, Montr\'{e}al, QC, H3A 2T8, Canada}

\author{Edward Wilson-Ewing}
\email{edward.wilson-ewing@unb.ca}
\affiliation{Department of Mathematics and Statistics,
University of New Brunswick, Fredericton, NB, E3B 5A3, Canada}

\date{\today}


\begin{abstract}

We show that the constraints which follow from the {\it Trans-Planckian Censorship Conjecture} for inflationary cosmology can be strengthened if the pre-inflationary period the universe was dominated by radiation. The resulting upper bound on the energy scale of inflation is $\eta \sim 10^4 ~ {\rm GeV}$, close to the scale accessible to accelerator experiments.

\end{abstract}



\maketitle


\section{Introduction} 

\label{sec:intro}

Recently, Bedroya and Vafa \cite{BV} put forwards the {\it Trans-Planckian Censorship Conjecture} (TCC) according to which no modes which had an initial wavelength smaller than the Planck length are permitted to exit the Hubble horizon during cosmological evolution. A motivation for this conjecture comes from an analogy with Penrose's cosmic censorship hypothesis \cite{Penrose} according to which time-like singularities must be hidden from external observers by horizons. This not only shields the external observer from having access to the singular surface, but also shields the external observer from the region into which the non-unitary evolution due to the time-like singularity can extend. The TCC can be viewed as a generalization of Penrose's hypothesis to also shield observers from Planck-scale physics, within the context of cosmology.

The expansion of the universe leads to an increase in the physical wavelengths of all fluctuation modes. In an effective field theory (EFT) analysis it is necessary to impose an ultraviolet cutoff on the modes at a fixed physical wavelength. But since the physical wavelengths of cosmological perturbation modes increase as the universe expands, there is a unitarity problem in the EFT description since the Hilbert space of the fluctuation modes must grow as a function of time \cite{Weiss}.  During inflation, fluctuations oscillate on sub-Hubble scales but freeze out, grow in amplitude, and classicalize on super-Hubble scales (see, e.g., \cite{MFB} for a detailed review of the theory of cosmological perturbations, and \cite{RHBfluctsRev} for an overview). The TCC states that fundamental physics will prohibit fluctuation modes which were at any time trans-Planckian (and hence outside of the realm of the EFT) from ever becoming classical and accessible to late-time observers, and shields late-time observers from the unitary problems of the EFT description.

The TCC does not lead to any constraints on the evolution in Standard Big Bang cosmology since in this case the fluctuation modes do not ever exit the Hubble radius (rather, it is super-horizon modes that re-enter the Hubble radius). On the other hand, for an inflationary universe scenario with an early phase of almost exponential expansion, the TCC leads to severe constraints%
\footnote{The fact that the accelerated expansion of space during inflation leads to a {\it trans-Planckian problem} for fluctuations was already pointed out in \cite{Jerome1}, and see \cite{Jerome2} for a review with references to other works on this problem.}~\cite{BBLV}.
Assuming that the expansion rate is almost constant during the period of inflation and assuming that the universe after inflation is given by the radiation phase of Standard Big Bang cosmology%
\footnote{Constraints which can be derived from the TCC when relaxing the condition of immediate radiation-domination after the end of inflation have been considered in \cite{MMPZ, Relax1}, and relaxed bounds when abandoning the assumption of an almost constant value of the expansion rate have been studied in \cite{MMPZ, Relax2}---note that the post-inflation $w = - 1/3$ phase in the setup of \cite{MMPZ} has a space-time diagram which is identical to the limiting value of power-law inflation studied in \cite{Relax2} when the inflationary phase has a value of $w$ which approaches the value $-1/3$ from below (and in \cite{Berera3} in the context of warm inflation \cite{warm}), while TCC constraints on the shape of the inflaton potential in single-field slow-roll inflation are derived in \cite{shape}. In deriving the bound on $r$, it was assumed that the initial state of the fluctuations is the usual Minkowski-like vacuum. What happens when this assumption is relaxed has been considered in \cite{Relax3}.},
there is an upper bound on the energy scale of inflation~\cite{BBLV}
\be \label{Vbound}
V_{e} \, < \, 10^{10} ~ {\rm GeV} \, ,
\ee
which leads to an upper bound on the amplitude of the spectrum of primordial gravitational waves. This can be expressed as an upper bound on the tensor-to-scalar ratio $r$,
\be
r \, < \, 10^{-30}.
\ee
Note that these bounds are independent of any assumptions about single field or slow-roll inflation~\cite{BBLV}.

These bounds are minimal in the sense that they do not make any assumptions about the pre-inflationary evolution. Working in the context of Einstein gravity coupled to scalar field matter (the scalar field responsible for inflation), it is rather likely that there was a pre-inflationary phase of radiation-domination. As we show in this note, if this is the case then the constraints on the energy scale of inflation become more severe; see also Sec.~IV in \cite{MMPZ} for a strengthened TCC bound when including a pre-inflationary era.  In this paper, we pay particular attention to the case where the pre-inflationary phase is radiation-dominated, in which case it is possible to relate the Hubble rate with the temperature of the radiation fluid; this provides a clearer understanding of the TCC and its implications for early universe cosmology.  We also include the effect of the late-time epoch of matter domination, which was neglected in \cite{MMPZ}.  In particular, we find that if the pre-inflationary phase is radiation-dominated then the TCC implies that the energy scale of inflation is constrained to be $\eta \sim 10^4 ~ {\rm GeV}$, close to the scale accessible to accelerator experiments.

In the following we work in the context of a spatially flat Friedman-Lema{\^i}tre-Robertson-Walker cosmology.  The metric, expressed in terms of the cosmic time $t$ and the comoving coordinates ${\bf{x}}$, is
\be
\dd s^2 = -\dd t^2 + a(t)^2 \, \dd \mathbf{x}^2,
\ee
where the scale factor $a(t)$ depends only on $t$, and the Hubble expansion rate is given by $H(t) = \dot{a}/a$, with the dot denoting a derivative with respect to $t$.  We adopt natural units in which the speed of light $c = 1$ and denote by $m_{\rm Pl}$ the reduced Planck mass.

\section{Strengthening the TCC Bound for Inflationary Cosmology}

Figure 1 shows a space-time sketch of an inflationary cosmology with a pre-inflationary radiation phase. The vertical axis is time, the horizontal axis is the physical length scale. The inflationary phase (assuming constant $H$ during inflation) lasts from $t = t_i$ to $t = t_R$. The time $t_{p}$ is the Planck time when the background energy density reaches the Planck scale, and $t_0$ denotes the present time. The solid curve denoted by $l_H$ is the Hubble horizon
\be
l_H(t) \, = \, H^{-1}(t) \, ,
\ee
and the dashed line labelled by $k$ shows the physical length of a fluctuation mode which is entering the Hubble radius today. The $x$ value of the vertical dashed line denotes the Planck length.

\begin{figure*}[t]
\begin{center}
\includegraphics[scale=1.2]{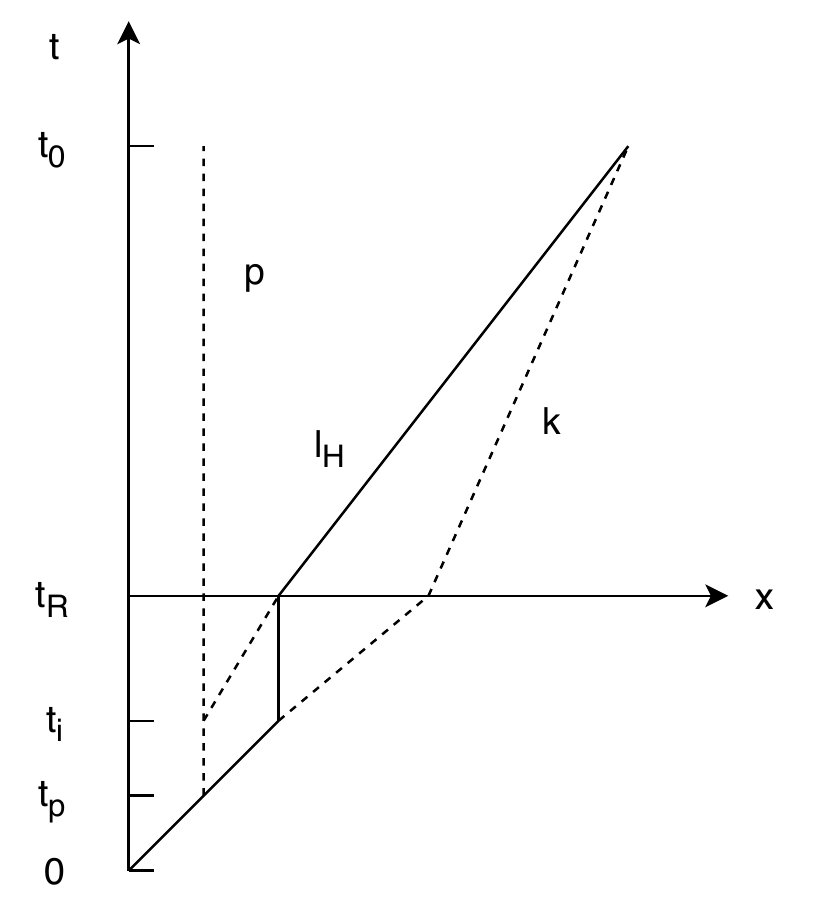}
\caption{Space-time sketch of the cosmology we are studying: an inflationary phase lasting from $t = t_i$ to $t = t_R$ with a preceding radiation phase from the Planck time $t_p$ to $t_i$. The horizontal axis denotes physical length, and the vertical axis is time. The solid line is the Hubble radius $l_H$, the vertical dashed line labelled $p$ indicates the Planck length, the dashed line labelled by $k$ represents the physical length of a fixed comoving scale which enters the Hubble radius at the present time $t_0$. In order for inflation to provide a causal mechanism for structure formation, this scale has to be sub-Hubble at the beginning of inflation, a condition which in this sketch is shown to be saturated.}
\end{center}
\end{figure*}

The minimal constraint on inflationary cosmology resulting from the TCC originates by demanding that no scale which has length smaller than the Planck length $l_{\rm Pl} = m_{\rm Pl}^{-1}$ at the beginning of inflation ever exits the Hubble radius,
\be \label{TCC}
\frac{a(t_R)}{a(t_i)} \, l_{\rm Pl} \, \leq \, H^{-1}(t_R) \, .
\ee
In the sketch of Figure 1 this condition is saturated: the dashed line starting with Planck length at the beginning of inflation just barely reaches the Hubble length at the end of inflation.

As is clear from Figure 1, in a cosmology where the universe is expanding during the pre-inflationary phase, if the minimal TCC condition \eqref{TCC} is saturated then there are modes that initially have a physical wavelength equal to or even shorter than $l_{\rm Pl}$ at a time between $t_p$ (before which the EFT description of the background space-time breaks down) and $t_i$ which exit the Hubble radius during the inflationary period before $t_R$.  These trans-Planckian modes correspond to curves which are displaced parallel downwards from the dashed curve which reaches the Hubble radius at $t_R$.  Of course, this possibility would violate the TCC.  Therefore, in an expanding universe the TCC also requires the stronger condition that
\be \label{TCC2}
\frac{a(t_R)}{a(t_p)} \, l_{\rm Pl} \, \leq \, H^{-1}(t_R) \, .
\ee
Assuming that the pre-inflationary phase is radiation-dominated, the slope of the dashed curve is steeper in the radiation phase than the slope of the Hubble radius curve%
\footnote{Note that the slope of the dashed curve is steeper than the slope of the Hubble radius for all (non-inflationary) matter fields with an equation of state $\omega = p/\rho > -1/3$.  The smaller the equation of state is, the stronger the new condition \eqref{TCC2} is on the Hubble radius during inflation.};
the former scales as $a(t) \sim t^{1/2}$, the latter as $t$.  Due to the growth in $a(t)$ during the pre-inflationary phase, the TCC now implies a stronger bound on the energy scale of inflation by a factor of $a(t_i)/a(t_p)$.

In order for inflation to provide the possibility of a causal mechanism to explain the observed structure of the universe on cosmological scales, the comoving scale corresponding to the current Hubble radius (denoted by $k$) must originate inside the Hubble radius at the beginning of inflation
\be \label{SF}
H_0^{-1} e^{-N} \frac{T_0}{T_R} \, \leq \, H^{-1}(t_R) \, .
\ee
Here $N$ is the number of e-foldings of inflation, $H_0$ denotes the current Hubble constant, $T_0$ the current temperature of the cosmic microwave background, and $T_R$ the temperature at the end of inflation. The condition (\ref{SF}) assumes that the radiation phase of Standard Big Bang cosmology starts right after inflation, and that the entropy production between $t_R$ and $t_0$ is negligible%
\footnote{Here and in the following we neglect factors of order 1 in the analysis.}
(in which case the ratio of scale factors is the inverse of the ratio of the temperatures).

Since in a radiation-dominated space-time $T \propto a^{-1}$, the condition (\ref{TCC2}) can be rewritten in terms of temperatures,
\be \label{TCC3}
\frac{T_p}{T_i} e^N l_{\rm Pl} \, \leq \, H^{-1}(t_R) \, ,
\ee
where $T_i$ and $T_p$ are the radiation temperatures at times $t_i$ and $t_p$, respectively.

The lower bound (\ref{SF}) and upper bound (\ref{TCC3}) on the duration of inflation $e^N$ are compatible provided that
\be \label{cond}
\frac{H(t_R)}{H_0} \frac{T_0}{T_R} \, \leq \, \frac{1}{l_{\rm Pl} H(t_i)} \frac{T_i}{T_p} \, ,
\ee
using the assumption that $H(t_i) = H(t_R)$.  The Friedman equation for a thermal bath of radiation relates the Hubble rate with the temperature,
\be
H(T)^2 = \frac{1}{3 m_{\rm Pl}^2} ~ g^*(T) ~ T^4 \, ,
\ee
where $g^*$ is the number of spin degrees of freedom in the thermal bath. Assuming that
$g^*$ is of order unity between $t_p$ and $t_i$ and also between $t_R$ and $t_0$,
condition (\ref{cond}) becomes
\be \label{bound}
T_i T_R \, \leq \, \frac{m_{\rm Pl}^2}{T_0} H_0 \, ,
\ee
using also $T_p = m_{\rm Pl} = l_{\rm Pl}^{-1}$.

The Friedman equation also implies
\be \label{Hzero}
H_0^2 \, = \, \frac{T_0^3 ~ T_{eq}}{3 \, m_{\rm Pl}^2} \, ,
\ee
where $T_{eq}$ is the temperature at the time of equal matter and radiation.  This relation holds since $H_{eq}^2 \propto T_{eq}^4$, and for $T < T_{eq}$ the universe is matter-dominated so $\rho \propto a^{-3} \propto (T/T_{eq})^3$.

Inserting (\ref{Hzero}) into (\ref{bound}) and assuming $T_i = T_R$ (no decrease in $H$ during inflation, and instantaneous reheating) then gives
\be
T_R \, \leq \, \sqrt{m_{\rm Pl}} \bigl( T_{eq} T_0 \bigr)^{1/4} \, \sim \, 10^4 ~ {\rm{GeV}} \, ,
\ee
up to factors of order 1.  Note that this scale is not too far above the reach of current accelerator experiments.

Since the amplitude ${\cal P}(h)$ of the spectrum of primordial gravitational waves produced during inflation is given by \cite{Starob}
\be
{\cal{P}}(h) \, \sim \, \left( \frac{H}{m_{\rm Pl}} \right)^2 \, \sim \, \left( \frac{T_R}{m_{\rm Pl}} \right)^4 \, < \, 10^{-56} \, ,
\ee
the bound on the tensor to scalar ratio $r$ becomes (given the observed amplitude of the scalar spectrum ${\cal P(R)} \sim 10^{-9}$)
\be
r \, < \, 10^{-47} \, ,
\ee
which is many orders of magnitude smaller than the amplitude of stochastic gravitational waves produced by other processes between the end of inflation and the present time.

\section{Conclusions and Discussion} \label{conclusion}

In this note we have shown that the upper bound on the energy scale of inflation tightens substantially compared to what was obtained in \cite{BBLV} if the universe is radiation-dominated (and expanding) between the Planck time and the onset of inflation. In fact, the bound reduces to an energy scale only slightly higher than the scale being probed by terrestrial accelerator experiments.  In this case the lowered upper bound on the energy scale of inflation implies an even stronger bound on the amplitude of the spectrum of gravitational waves.  More generally, any pre-inflationary era during which the universe expands will strengthen TCC bounds on the energy scale of inflation.  For a pre-inflationary expanding phase with a constant equation of state $\omega$ (by definition, in the pre-inflationary phase $\omega > -1/3$), the strengthened TCC bound is
\be
\left( \frac{T_R}{m_{\rm Pl}} \right)^{3 \, - \, \tfrac{4}{3(1+\omega)}} \, \leq \, \sqrt \frac{ T_{eq} T_0 }{m_{\rm Pl}^2} \, .
\ee
A similar strengthened bound (although neglecting the late-time era of matter-domination) was derived in \cite{MMPZ}.  We see that the upper bound on $T_R$ decreases and becomes stronger if $\omega$ decreases. 

Note that the smaller the energy scale of inflation, the more contrived it becomes to construct an inflationary model which yields the observed amplitude of the spectrum of cosmological perturbations.  For example, for slow-roll single-field inflation, the TCC bound on the tensor-to-scalar ratio $r$ implies that the slow-roll parameter $\epsilon = - \dot{H} / H^2 \leq 10^{-48}$.  However, these results (and also those of \cite{BBLV}) can be relaxed by dropping one of the various assumptions we have made: an expanding radiation-dominated pre-inflationary phase, standard cosmology after inflation, an almost constant value of $H$ during inflation, a number of spin degrees of freedom in the thermal bath of the radiation fluid $g^*$ of order unity, and the standard mechanism for producing the observed nearly scale-invariant fluctuations. But on the other hand, relaxing any one of these assumptions implies departing from the simplicity of the original inflationary scenario.

Perhaps the simplest departure is to assume that the pre-inflationary phase is contracting rather than expanding, with a cosmic bounce occurring near the onset of inflation \cite{mod2}; inflationary models like these are not subject to the strengthened TCC constraints derived here.

Finally, recall that an important motivation for the TCC is to ensure that inflation can be described in an EFT framework.  On the other hand, a better understanding of fundamental physics could remove this motivation by providing a UV-complete description of inflation.  In this case, the {\it trans-Planckian problem} \cite{Jerome1} of fluctuations becomes a {\it trans-Planckian window of opportunity} to test fundamental physics---whether string theory, loop quantum gravity, asymptotic safety, or other theories---using cosmological observations.  Our best candidate for a theory that unifies all forces of nature and is complete in the ultraviolet is superstring theory. In the context of string theory, there have been a series of arguments which indicate that inflation is hard to realize in string theory \cite{noinflation}. They go under the name {\it Swampland Constraints} \cite{Vafa} (see \cite{Vafa-rev} for recent reviews, and see \cite{Samuel} for a recent analysis from the point of view of {\it string gas cosmology} \cite{BrV}). Thus, from the perspective of string theory one cannot simply discard conclusions coming from the TCC by invoking quantum gravity effects. In fact, as studied in \cite{BV} and \cite{Suddho}, there is a close connection between the swampland conditions and the TCC. In particular, both exclude the presence of a stable vacuum with a positive cosmological constant, but both are consistent with a phase of dark energy domination of cosmology at the present time (see, e.g., \cite{Lavinia}). 

$~$

\acknowledgments


E.W.-E.~thanks McGill University for hospitality during the completion of this work.

R.B.~is supported in part by the Natural Science and Engineering Research Council of Canada and by the Canada Research Chair program, and E.W.-E.~is supported in part by the Natural Science and Engineering Research Council of Canada and by a Harrison McCain Foundation Young Scholars Award.


\begin{thebibliography}{99}



\bibitem{BV}
A.~Bedroya and C.~Vafa,
  ``Trans-Planckian Censorship and the Swampland,''
  arXiv:1909.11063 [hep-th].

\bibitem{Penrose}
R.~Penrose,
  ``Gravitational collapse: The role of general relativity,''
  Riv.\ Nuovo Cim.\  {\bf 1}, 252 (1969)
  [Gen.\ Rel.\ Grav.\  {\bf 34}, 1141 (2002)].

\bibitem{Weiss}
N.~Weiss,
  ``Constraints on Hamiltonian Lattice Formulations of Field Theories in an Expanding Universe,''
  Phys.\ Rev.\ D {\bf 32}, 3228 (1985).
  doi:10.1103/PhysRevD.32.3228

\bibitem{MFB}
 V.F. Mukhanov, H.A. Feldman and R.H. Brandenberger,
 ``Theory of Cosmological Perturbations''
 Physics Reports \textbf{215}, 203 (1992).

\bibitem{RHBfluctsRev}
R.~H.~Brandenberger,
 ``Lectures on the theory of cosmological perturbations,''
 Lect.\ Notes Phys.\  {\bf 646}, 127 (2004)
 doi:10.1007/978-3-540-40918-25
 [hep-th/0306071].

\bibitem{BBLV}
A.~Bedroya, R.~Brandenberger, M.~Loverde and C.~Vafa,
  ``Trans-Planckian Censorship and Inflationary Cosmology,''
  arXiv:1909.11106 [hep-th].

\bibitem{Jerome1}
J.~Martin and R.~H.~Brandenberger,
  ``The Trans-Planckian problem of inflationary cosmology,''
  Phys.\ Rev.\ D {\bf 63}, 123501 (2001)
  doi:10.1103/PhysRevD.63.123501
  [hep-th/0005209].

\bibitem{Jerome2}
R.~H.~Brandenberger and J.~Martin,
  ``Trans-Planckian Issues for Inflationary Cosmology,''
  Class.\ Quant.\ Grav.\  {\bf 30}, 113001 (2013)
  doi:10.1088/0264-9381/30/11/113001
  [arXiv:1211.6753 [astro-ph.CO]].
  
\bibitem{MMPZ}
S.~Mizuno, S.~Mukohyama, S.~Pi and Y.~L.~Zhang,
  ``Universal Upper Bound on the Inflationary Energy Scale from the Trans-Planckian Censorship Conjecture,''
  arXiv:1910.02979 [astro-ph.CO].

\bibitem{Relax1}
M.~Dhuria and G.~Goswami,
  ``Trans-Planckian Censorship Conjecture and Non-thermal post-inflationary history,''
  arXiv:1910.06233 [astro-ph.CO];\\
  M.~Torabian,
  ``Non-Standard Cosmological Models and the trans-Planckian Censorship Conjecture,''
  arXiv:1910.06867 [hep-th];\\
 H.~H.~Li, G.~Ye, Y.~Cai and Y.~S.~Piao,
  ``Trans-Planckian censorship of multi-stage inflation and dark energy,''
  arXiv:1911.06148 [gr-qc];\\
M.~Torabian,
  ``Breathing Comoving Hubble: Initial Condition and Eternity in view of the trans-Planckian Censorship Conjecture,''
  arXiv:1911.12304 [hep-th].
  
\bibitem{Relax2}
V. Kamali and R. Brandenberger, 
``Relaxing the TCC Bound on Inflationary Cosmology?''
arXiv:2001.00040 [hep-th].

\bibitem{Berera3}   
S.~Das,
  ``Distance, de Sitter and Trans-Planckian Censorship conjectures: the status quo of Warm Inflation,''
  arXiv:1910.02147 [hep-th];\\
V.~Kamali, M.~Motaharfar and R.~O.~Ramos,
  ``Warm brane inflation with an exponential potential: a consistent realization away from the swampland,''
  arXiv:1910.06796 [gr-qc];\\
A.~Berera and J.~R.~Calderon,
  ``Trans-Planckian censorship and other swampland bothers addressed in warm inflation,''
  arXiv:1910.10516 [hep-ph];\\
 G.~Goswami and C.~Krishnan,
  ``Swampland, Axions and Minimal Warm Inflation,''
  arXiv:1911.00323 [hep-th].

\bibitem{warm}
A.~Berera,
  ``Warm inflation,''
  Phys.\ Rev.\ Lett.\  {\bf 75}, 3218 (1995)
  doi:10.1103/PhysRevLett.75.3218
  [astro-ph/9509049].

\bibitem{shape}
K.~Kadota, C.~S.~Shin, T.~Terada and G.~Tumurtushaa,
  ``Trans-Planckian censorship and single-field inflaton potential,''
  arXiv:1910.09460 [hep-th].

\bibitem{Relax3}   
 S.~Brahma,
  ``Trans-Planckian censorship, inflation and excited initial states for perturbations,''
  arXiv:1910.04741 [hep-th].

\bibitem{Starob}
 A.~A.~Starobinsky,
  ``Spectrum of relict gravitational radiation and the early state of the universe,''
  JETP Lett.\  {\bf 30}, 682 (1979)
  [Pisma Zh.\ Eksp.\ Teor.\ Fiz.\  {\bf 30}, 719 (1979)].

\bibitem{mod2}
I.~Agullo, A.~Ashtekar and W.~Nelson,
  ``A Quantum Gravity Extension of the Inflationary Scenario,''
  Phys.\ Rev.\ Lett.\  {\bf 109}, 251301 (2012)
  doi:10.1103/PhysRevLett.109.251301
  [arXiv:1209.1609 [gr-qc]]; \\
Y.~Cai and Y.~S.~Piao,
  ``Pre-inflation and Trans-Planckian Censorship,''
  arXiv:1909.12719 [gr-qc].

\bibitem{noinflation}
P.~Agrawal, G.~Obied, P.~J.~Steinhardt and C.~Vafa,
  ``On the Cosmological Implications of the String Swampland,''
  Phys.\ Lett.\ B {\bf 784}, 271 (2018)
  doi:10.1016/j.physletb.2018.07.040
  [arXiv:1806.09718 [hep-th]].

\bibitem{Vafa}
H.~Ooguri and C.~Vafa,
 ``On the Geometry of the String Landscape and the Swampland,''
 Nucl.\ Phys.\ B {\bf 766}, 21 (2007)
 doi:10.1016/j.nuclphysb.2006.10.033
 [hep-th/0605264];\\
G.~Obied, H.~Ooguri, L.~Spodyneiko and C.~Vafa,
 ``De Sitter Space and the Swampland,''
 arXiv:1806.08362 [hep-th].

\bibitem{Vafa-rev}
T.~D.~Brennan, F.~Carta and C.~Vafa,
 ``The String Landscape, the Swampland, and the Missing Corner,''
 PoS TASI {\bf 2017}, 015 (2017)
 doi:10.22323/1.305.0015
 [arXiv:1711.00864 [hep-th]];\\
E.~Palti,
 ``The Swampland: Introduction and Review,''
 arXiv:1903.06239 [hep-th].

\bibitem{Samuel}
S.~Laliberte and R.~Brandenberger,
  ``String Gases and the Swampland,''
  arXiv:1911.00199 [hep-th].

\bibitem{BrV}
R.~H.~Brandenberger and C.~Vafa,
  ``Superstrings in the Early Universe,''
  Nucl.\ Phys.\ B {\bf 316}, 391 (1989).
  doi:10.1016/0550-3213(89)90037-0

\bibitem{Suddho}
P.~Draper and S.~Farkas,
  ``Transplanckian Censorship and the Local Swampland Distance Conjecture,''
  arXiv:1910.04804 [hep-th];\\
S.~Brahma,
  ``The trans-Planckian censorship conjecture from the swampland distance conjecture,''
  arXiv:1910.12352 [hep-th];\\
R.~Saito, S.~Shirai and M.~Yamazaki,
  ``Is Trans-Planckian Censorship a Swampland Conjecture?,''
  arXiv:1911.10445 [hep-th].

\bibitem{Lavinia}
L.~Heisenberg, M.~Bartelmann, R.~Brandenberger and A.~Refregier,
  ``Dark Energy in the Swampland,''
  Phys.\ Rev.\ D {\bf 98}, no. 12, 123502 (2018)
  doi:10.1103/PhysRevD.98.123502
  [arXiv:1808.02877 [astro-ph.CO]].

                                                       

\end{thebibliography}
\end{document}